\documentclass[a4paper,12pt]{article}

\usepackage{amsmath,amssymb}
\usepackage[normalem]{ulem}
\usepackage{graphicx}
\renewcommand{\baselinestretch}{1.0}

\newcommand{\be}{\begin{equation}}
\newcommand{\ee}{\end{equation}}

\def\aprge{\buildrel > \over {_{\sim}}}
\newcommand{\bea}{\begin{eqnarray}}
\newcommand{\eea}{\end{eqnarray}}
\newcommand{\ba}{\begin{array}}
\newcommand{\ea}{\end{array}}
\newcommand{\balg}{\begin{align}}
\newcommand{\ealg}{\end{align}}
\newcommand{\bit}{\begin{itemize}}
\newcommand{\eit}{\end{itemize}}
\newcommand{\trm}[1]{\textrm{#1}}

\newcommand{\Mpc}{\trm{\Mpc}}
\newcommand{\yr}{\trm{\yr}}
\newcommand{\eV}{\trm{\eV}}

\begin{document}
\topmargin 0pt
\oddsidemargin=-0.4truecm
\evensidemargin=-0.4truecm
\renewcommand{\thefootnote}{\fnsymbol{footnote}}

\newpage
\setcounter{page}{1}
\vspace*{-2.0cm}
\begin{flushright}
\vspace*{-0.2cm}

\end{flushright}
\vspace*{0.5cm}

\begin{center}
{\Large \bf Active to sterile neutrino oscillations: \\ 

\vspace{.2cm}
Coherence and MINOS results
}
\vspace{0.5cm}

{D. Hernandez and A. Yu. Smirnov\\
\vspace*{0.2cm}
{\it The Abdus Salam International Centre for Theoretical Physics,\\
Strada Costiera 11, I-34013 Trieste, Italy }
}
\end{center}


\renewcommand{\thefootnote}{\arabic{footnote}}
\setcounter{footnote}{0}
\renewcommand{\baselinestretch}{0.9}

\begin{abstract}

We study the $\nu_\mu - \nu_s$ oscillation effects in the near detector of the 
MINOS experiment. Conceptually, the MINOS search for sterile neutrinos 
with mass $\sim 1$ eV realizes an interesting  situation of partial decoherence of the neutrino state at the 
production. This corresponds to a difference of  energies of the two mass eigenstates that is comparable with or 
bigger than the width of the initial state (pion). We show that these effects modify 
the MINOS bound on mixing of sterile neutrino for 
$\Delta m^2_{41} \aprge 0.5$ eV$^2$ and make the experiment insensitive to 
oscillations with $\Delta m^2_{41} \gtrsim 15$ eV$^2$. Oscillations with 
$\Delta m^2_{41} =  (1 - 3)$ eV$^2$  could  explain some deficit of events  
observed in the low energy bins in the near detector and correspondingly the 
excess of events in the far detector. 

\end{abstract}

\section{Introduction}

Recently the MINOS collaboration has published a  very stringent limit on the active to sterile neutrino mixing 
from neutral-current 
interactions \cite{Adamson:2011ku}.  The  $\nu_\mu - \nu_s$ mixing angle, 
$\theta_{24}$, should be smaller than  
\be 
\theta_{24} < 7^{\circ} ~~( 90 \% ~~{\rm C.L.})~~ {\rm for} ~~m_4 \gg 
m_3   
\label{limit}
\ee
at 1-3 mixing value, $\theta_{13} = 0$. 
For maximally allowed $\theta_{13}$ 
the limit becomes slightly weaker: $\theta_{24} < 8^{\circ}$. 
Eq.~(\ref{limit}) corresponds  to $|U_{\mu 4}|^2 \leq 0.015$. Being valid, as 
we will confirm, for $\Delta m^2 < 0.5$ eV$^2$ this limit essentially 
excludes the sterile neutrino  interpretation of the LSND~\cite{lsnd} and
MiniBooNE~\cite{miniboone} results. Indeed, the required mixing for 
their explanation increases when $\Delta m^2$ decreases, whereas 
the bounds on $|U_{\mu 4}|^2$  from MINOS and $|U_{e 4}|^2$ from reactor 
experiments  change weakly with $\Delta m^2$ in the range of interest.    

The limit in Eq.~\eqref{limit}, however, can not be applied to oscillations 
with  $\Delta m^2 \sim 1$ eV$^2$ relevant for  LSND and 
MiniBooNE. Indeed, the analysis in \cite{Adamson:2011ku} has been performed 
assuming that no oscillation-induced change of the neutrino event rate is 
measurable in  the near detector (ND). 
The ND is located at the distance $L = 1.04$ km from the beam target, 
and therefore the oscillations with $\Delta m^2 \sim 1$ eV$^2$ 
can not be neglected. If for instance,  the neutrino energy equals $E_\nu = 1$ GeV, the 
oscillation length is 2.5 km. 
The first oscillation minimum is at $E_\nu = 0.8$ GeV~\footnote{The possibility of a small oscillation 
effect in the ND in MINOS for $\Delta m^2\sim 1 \trm{eV}^2$  was  mentioned previously in 
\cite{Meloni:2010zr}}.   

Admittedly, the systematic errors in the ND are bigger than the oscillation effect and, therefore, the latter is not ``measurable'' by this detector alone. However, the search of oscillations and  the MINOS bound on the angle $\theta_{24}$ are based on the comparison of the signals in the ND and far detector (FD). Namely, the  energy spectrum of the neutral current (NC) events in the  ND is extrapolated to the FD and  confronted with the data. This allows one to substantially reduce the systematic errors.  It is the difference of the oscillation effects in the ND and FD that allows one to obtain the bound on the oscillation parameters. Therefore, oscillation effects in the ND can not be neglected when ``propagating'' oscillation predictions to the FD.

The search for neutrino oscillations from free pion decay with $\Delta m^2_{42} \sim 1$ eV$^2$ realizes the conceptually interesting situation in which
the coherence of the neutrino state is partially broken at the production.
Recall that in a given experimental setup, coherence is destroyed at production if it is possible to identify in principle (using  kinematics of the process) which mass eigenstate is produced (see \cite{akhm} and refernces therein). 

For free pion decay and undetected muon the energy uncertainty, $\sigma_E$, is determined by the  pion decay  rate $\sigma_E \sim  \Gamma_\pi$. The difference of energies of the mass states equals $\sim  \Delta m^2/2E$ and therefore the parameter
\be
\xi \equiv \frac{\Delta m^2}{2E \Gamma} = 2\pi \frac{l_{dec}}{l_{\nu}}
\label{eq:par}
\ee
can be considered as a measure of the decoherence of the neutrino state. We will call it the {\it decoherence parameter}. Here $l_{dec} = 1/\Gamma$ 
is the decay length and $l_{\nu} = 4\pi E/\Delta m^2$ is the oscillation length. If  $\xi 
\gg 1$  the two mass states can
be resolved  whereas for $\xi \ll 1$ the uncertanty is large and
decoherence can be neglected. 
In the case of MINOS, due to the large size of the pipe-line ($l_p = 675$ m), 
pions undergo free decay if collisions are neglected. 
For instance, for $E_\pi \sim 10$ GeV the decay length, 
$l_{dec} = c \tau_0 \gamma_\pi \sim 560$ m, is smaller than $l_p$. 
If  $\Delta m^2 \sim 1 \trm{ eV}^2$, one has $\xi \sim 1$ and decoherence at the production can not be neglected.
This is the case for most of the pion spectrum in MINOS, namely, the size of the region of coherent neutrino production is given by the decay length and is comparable with the oscillation length.
For larger energies of the pion the decay region is determined by the size of pipe.

In this paper we show that partial decoherence  of the neutrino state at the production modifies the standard  oscillation result in the MINOS experiment. We find the spectra of events in the ND and in the FD with $\nu_\mu - \nu_s$ 
oscillations taken into account. We estimate the modification of the bounds on sterile neutrino mixing when oscillations at the ND are included. We notice that  the deficit of the events in the low energy bins in the ND and some excess of events in the FD could be explained  by oscillations  into sterile neutrinos. We 
elaborate on the coherence condition in other experiments searching for sterile neutrinos with masses around 1 eV (MiniBooNE,  new proposed experiments) to which similar considerations can be applied.

The paper is organized as follows. In Sec.~2 we consider  the oscillation effects in the 
MINOS ND assuming that neutrinos are produced incoherently. In Sec.~3 we discuss the 
loss of coherence at  the production and compute the effective oscillation probability 
considering coherent neutrino production along the pion trajectory. 
We will show that 
in the MINOS setup, the same result as in the incoherent case can be obtained from the 
wave packet  consideration for free pion decay under some approximations. 
In Sec.~4 we compute the numbers of events in the MINOS ND and FD  
and discuss modifications of the 
bounds when oscillations in the ND are taken into account. Conclusions are presented in Sec.~5
 
\section{Oscillation effects in the near detector. The case of incoherent production}

In this section we compute the oscillation effect in the ND considering
incoherent production of neutrinos along the pion trajectory.
We find first the oscillation probability
for neutrinos produced in certain space-time point $(x_S, t_S)$
of the production region (decay pipe) and then integrate
this probability folded with the number of pions decaying
in the point  $(x_S, t_S)$  over the decay pipe.

Let us consider pions with energy $E_\pi$  
produced in the target at origin, $x=0$,  
and compute the neutrino flux obtained at a detector located at $x=L$
taking into account oscillations. 
The flux of neutrinos is given by
\be
F_\nu (E_\nu) = 
\int_0^{l_p} dx P_{\mu\mu} (E_\nu, L - x) F_\nu (E_\nu, x),  
\label{eq:flux}
\ee
where  $P_{\mu\mu}$ is the  $\nu_\mu - \nu_\mu$  survival probability and $l_p$  is the 
length of the decay pipe.  
$F_\nu (E_\nu, x) dx$ is the  density of neutrino flux  in the interval $(x,\, x + dx)$,  produced from pion 
decay at a distance $x$ from the target. 
After appropriate integration over the angular variables,  the neutrino flux is equal to 
\be
F_\nu (E_\nu, x) = \int_{E_\pi^{min}}^{\infty} d E_\pi  F_\pi (E_\pi)
e^{-\Gamma (E_\pi) x} \Gamma (E_\pi) K(E_\pi, E_\nu), 
\label{eq:numb}
\ee
where $F_\pi (E_\pi)$ is the flux of pions with energy 
$E_\pi$, $\Gamma = m_\pi \Gamma_0/ E_\pi$ is the decay 
rate in the laboratory frame,   $\Gamma_0 = 1/\tau_0 =  2.5 \cdot 10^{-8} ~{\rm eV}$ is 
the decay rate in the  pion rest frame; 
\be 
E_\pi^{min} \approx  \frac{E_\nu m_\pi}{2 E_0}
\ee 
is the minimal pion energy required to produce a neutrino with energy $E_\nu$. 
$E_0$  is the  neutrino energy in the rest frame of pion 
\be
E_0 = p_0 = \frac{m_\pi^2 - m_\mu^2}{2 m_\pi},   
\ee
where we have neglected the mass of the neutrino. 
$K(E_\pi, E_\nu)$ is the probability 
that a pion with energy $E_\pi$ emits a neutrino with energy $E_\nu$. 
Similarly, one could write the contribution to the neutrino flux from two-body $K$ decay just by substituting $m_\pi \rightarrow m_K$,  $\Gamma_\pi \rightarrow \Gamma_K$.
For illustration purposes we will only consider here  neutrinos from pion decay which, 
in fact, constitute by far the main component of  the neutrino beam. 

Notice that the integral in Eq.~(\ref{eq:numb}) is nothing but the integral of the oscillation probability multiplied by the density of pions over the production region, that is the incoherent  summation over neutrino sources. This is realized when the size of neutrino wavepackets  is much smaller than the typical size of the source, $\sigma_x \ll l_p$. 

In what follows for simplicity we assume that the energy of neutrinos detected in the ND is 
uniquely related to the pion energy: 
\be
E_\nu = \alpha E_\pi, 
\label{eq:approx}
\ee
where $\alpha = const$. This approximation allows us to  
elucidate the physics involved and still obtain rather precise results. 
We consider validity  of the approximation 
and determine value of  the parameter $\alpha$ in the Appendix. 
In the approximation of Eq.~(\ref{eq:approx}) 
\be
K(E_\pi, E_\nu) = \delta (E_\pi - \alpha^{-1} E_\nu )  
\label{eq:kkk}
\ee
and the integration in Eq.~(\ref{eq:numb}) is trivial giving 
\be
F_\nu (E_\nu, x) =   F_\pi (\alpha^{-1} E_\nu) e^{-\Gamma (E_\nu)  x} 
\Gamma (E_\nu), 
\label{eq:numb1}
\ee
where 
\be
\Gamma (E_\nu) = \frac{\alpha \Gamma_0 m_\pi}{E_\nu}. 
\ee
Integration of $F_\nu (E_\nu, x)$ over $x$,  which corresponds to setting  $P_{\mu \mu} 
= 1$ in Eq.~(\ref{eq:flux}), gives the neutrino flux at the ND without oscillations: 
\be
F_\nu^0 (E_\nu) = F_\pi (\alpha^{-1} E_\nu) 
\left[1 - e^{-\Gamma (E_\nu) l_p} \right]. 
\label{eq:fzero}
\ee
This relation determines the flux of pions, $F_\pi$,
in terms of the non-oscillating neutrino flux. Inserting this pion flux into 
Eq.~(\ref{eq:numb1}) we obtain 
\be
F_\nu (E_\nu, x) =   F_\nu^0 (E_\nu) 
\frac{\Gamma (E_\nu) e^{-\Gamma (E_\nu) x}}{1 - e^{-\Gamma (E_\nu) l_p}} ~. 
\ee
Then, according to Eq.~(\ref{eq:flux}), the $\nu_\mu-$flux in the presence 
of oscillations equal
\be
F_\nu (E_\nu) = \frac{F_\nu^0 (E_\nu)} {1 - e^{-\Gamma (E_\nu) l_p}} 
\int_0^{l_p} dx \Gamma (E_\nu) e^{-\Gamma(E_\nu) x} P_{\mu\mu} (E_\nu, L - x). 
\label{eq:flux2}
\ee
This expression has been obtained essentially in the factorization approximation: the oscillation probability is multiplied by the production probability and later it is multiplied by the detection probability (cross-section). 

The oscillation probability can be written as 
\be
P_{\mu\mu} (E_\nu, L - x) = \bar{P} + 
\frac{1}{2} \sin^2 2 \theta_{24} \cos [\xi \Gamma (L - x)], 
\label{eq:pmumu}
\ee
where 
\be
\bar{P} = 1 - \frac{1}{2} \sin^2 2 \theta_{24} 
\ee
is the averaged oscillation probability, and  the decoherence parameter $\xi$   
introduced in (\ref{eq:par}) can be rewritten as 
\be 
\xi \equiv  \frac{\Delta m^2_{42}}{2E_\nu \Gamma} 
=  \frac{\Delta m^2_{42}}{2 \alpha m_\pi \Gamma_0}.  
\label{eq:coheren}
\ee
Notice  that $\xi$  does not depend on 
the baseline or neutrino  energy but only on intrinsic 
parameters of the pion and the neutrinos.

Using the flux in Eq.~(\ref{eq:flux2}) we can introduce the ratio of fluxes with and 
without  oscillations which can be considered  as the effective survival probability. 
From  Eqs.~(\ref{eq:flux2}) and (\ref{eq:pmumu}) we obtain
\be
P_{eff} \equiv \frac{F_\nu }{F_\nu^0} =  \bar{P} + \frac{1}{2} \sin^2 2 \theta_{24} 
\frac{\Gamma}{{1 - e^{-\Gamma l_p}}}  \int_0^{l_p} dx  e^{-\Gamma  x} \cos [\xi \Gamma 
(L - x)], \label{eq:flux2a}
\ee
and the integration in Eq.~(\ref{eq:flux2a}) gives
\be
P_{eff} = 
\bar{P} + \frac{\sin^2 2\theta_{24}}{2(1 + \xi^2)} \cdot \frac{1}{1 - e^{-\Gamma l_p}}
\left[ \cos \phi_L + \xi \sin \phi_L  - 
e^{-\Gamma  l_p} \left( \cos(\phi_L - \phi_p) + \xi \sin(\phi_L - \phi_p) \right)\right].
\label{eq:fluxr}
\ee
Here  
\be
\phi_L \equiv  \frac{\Delta m^2}{2 E_\nu} L = \Gamma \xi L, ~~~~ 
\phi_p \equiv  \frac{\Delta m^2}{2 E_\nu} l_p = \Gamma \xi l_p  
\ee
are the oscillation phases acquired over the baseline $L$ and the pipe line, $l_p$. 

The most noticeable deviation of Eq.~(\ref{eq:fluxr}) 
from the standard formula for oscillations with the baseline $L$,   
\be 
P_{stand} =
\bar{P} + \frac{1}{2} \sin^2 2 \theta_{24} 
\cos \phi_L,  
\label{eq:stand}
\ee
is related to the factors which depend on  $\xi$ that lead to a suppression of the 
depth of oscillations and to a modification  of the oscillatory term. 
Eq.~\eqref{eq:fluxr} provides a clear interpretation  of the decoherence parameter. 
For very  short decay pipe, $\Gamma l_p \ll 1$, 
the standard  oscillation formula with 
baseline $L$, Eq.~\eqref{eq:stand},  which corresponds to a fully coherent  superposition of the 
mass eigenstates,  is recovered in the limit $\xi \ll 1$.  On the other hand, in the 
limit $\xi \rightarrow \infty$, $P_{eff}$ reduces to the  averaged probability which reflects  
complete decoherence. Therefore, indeed $\xi$ is a measure of  the decoherence introduced in 
the production process of the neutrino. The decoherence  effect thus disappears at  
large $\Gamma_0$ or small $\Delta m^2$. 
Other useful expression for $\xi$ is
\be
\xi = \frac{\phi_p}{ \Gamma l_p} 
\ee

In the limit of low energies,  $l_p \Gamma \gg 1$, we obtain    
\be
\frac{F_\nu }{F_\nu^0 } \approx
1 -  
\frac{\sin^2 2 \theta_{24}}{2(1 + \xi^2)}(\cos \phi_L + \xi \sin \phi_L),  
\label{eq:fluxr3}
\ee
and the oscillation pattern does not disappear in spite of the integration over 
large distances. The reason is that with  decrease of energy the oscillation length 
decreases as 
$l_\nu \propto E_\nu$ and the effective  region of the pion decay, $\Gamma  \propto 
E_\nu $  decrease in the same way. 

\begin{figure}[ht]
\begin{center}
\vskip 1cm
\includegraphics[width=8cm]{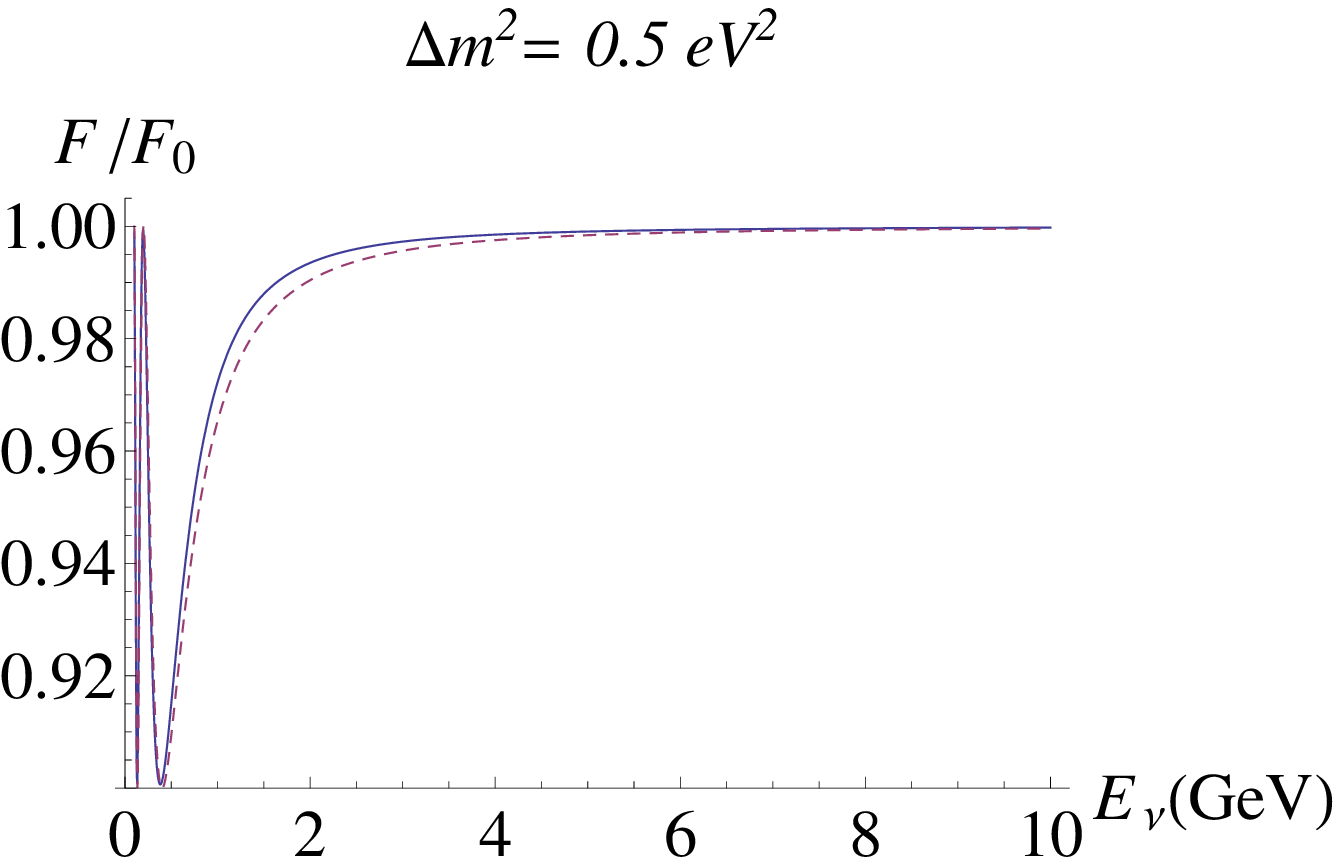} 
\includegraphics[width=8cm]{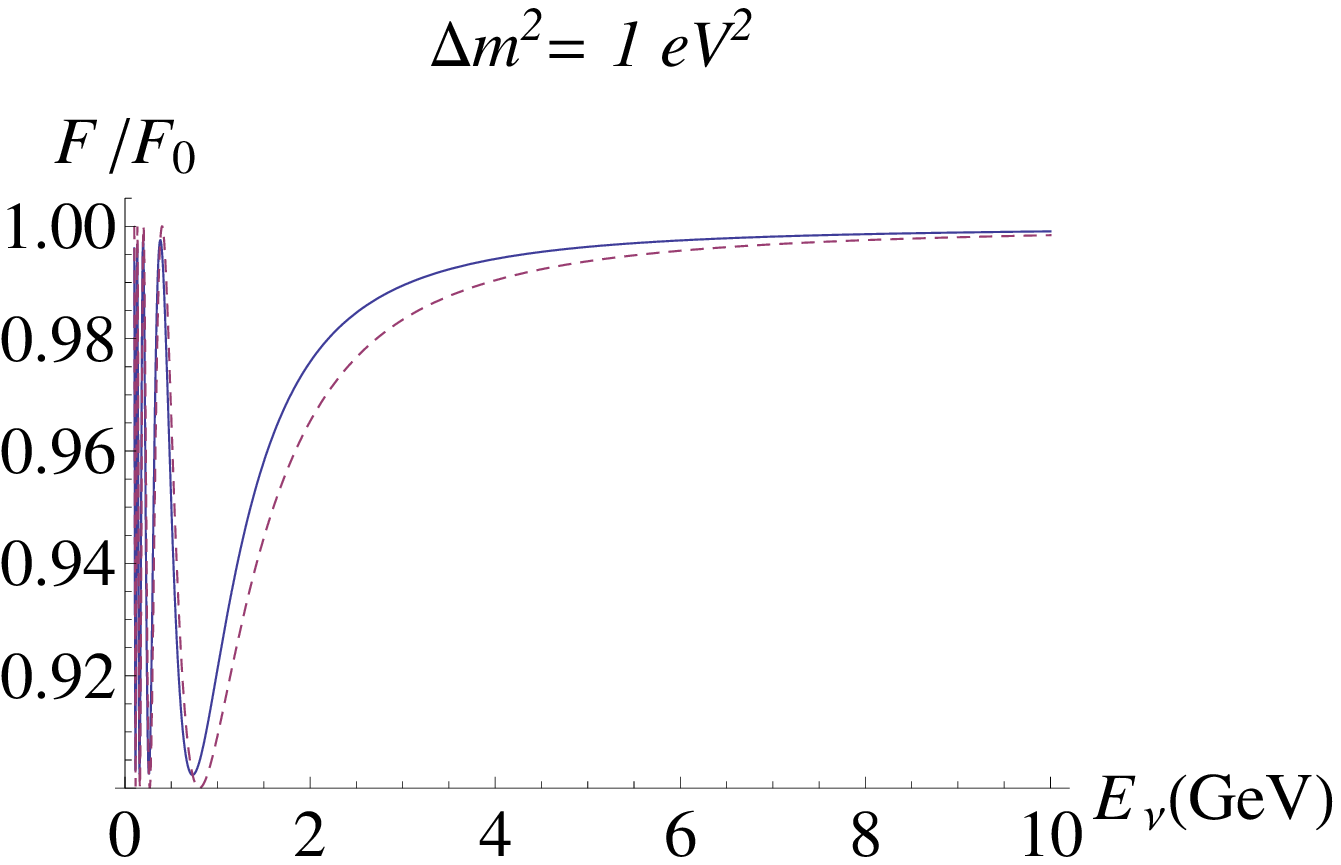}\\
\end{center}

\begin{center}
\includegraphics[width=8cm]{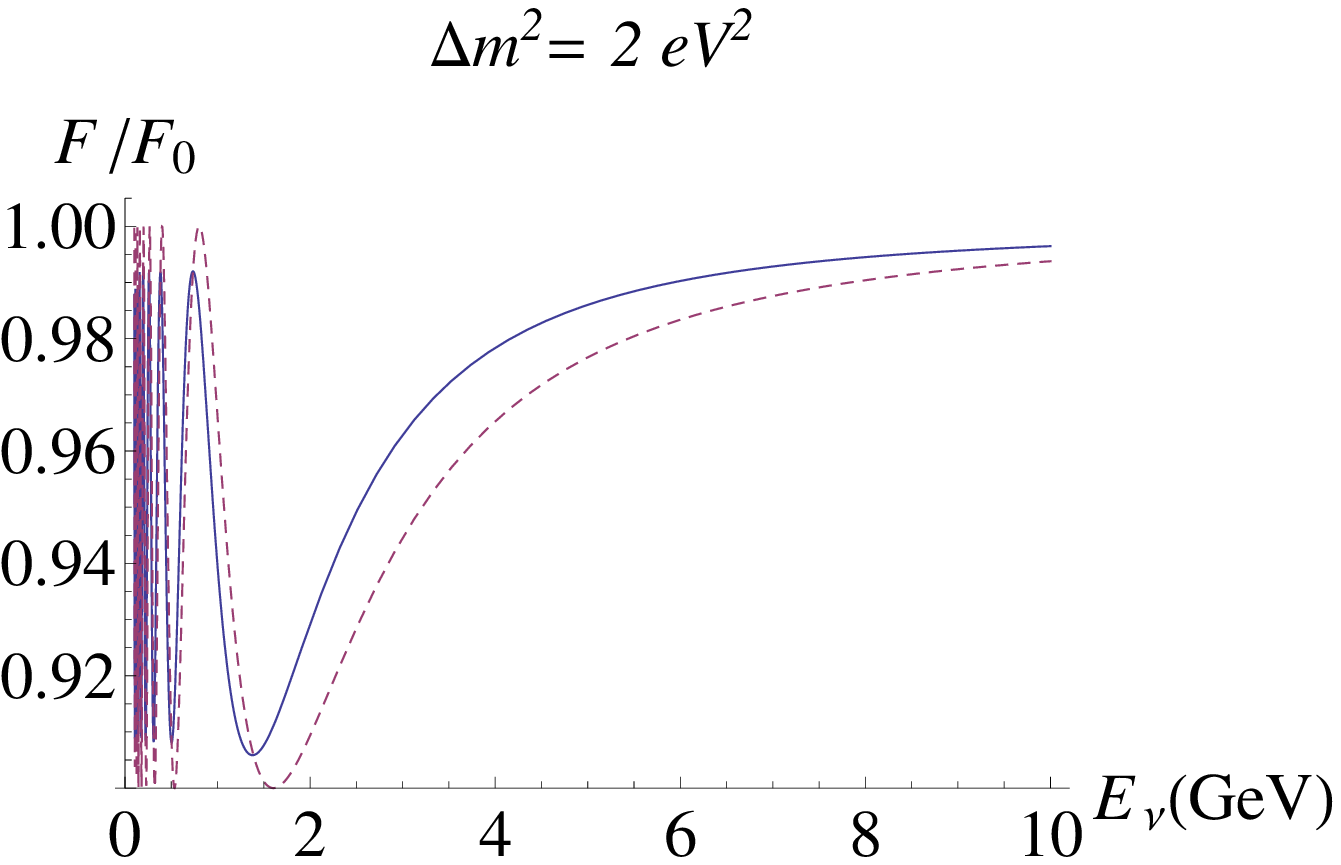} 
\includegraphics[width=8cm]{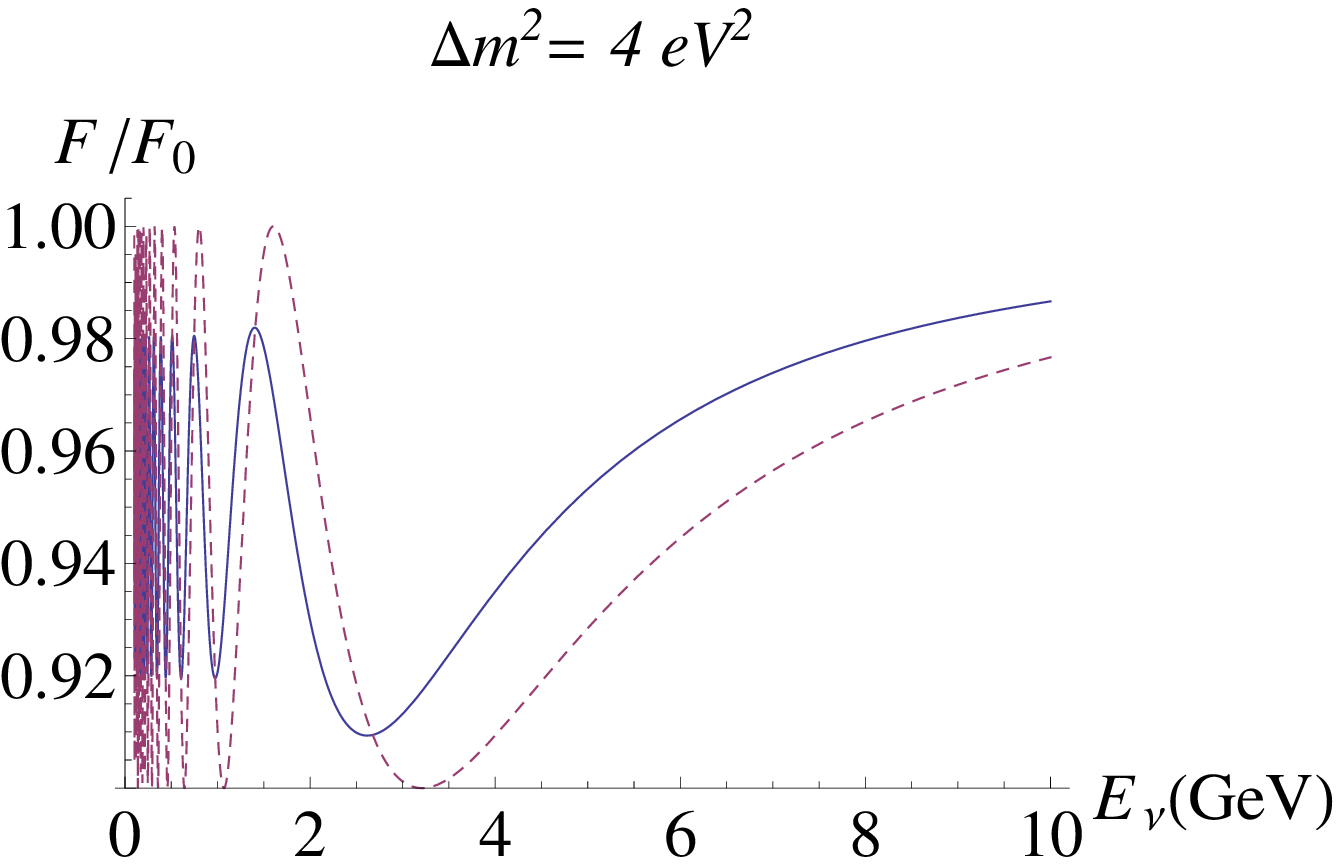}
\caption{The ratio of neutrino fluxes with and without oscillations 
as a function of neutrino energy for different values of $\Delta m^2_{42}$: 
0.5, 1 , 2, 4  eV$^2$, and 
$\sin^2 2\theta_{24} = 0.1$. For comparison the standard oscillation probability with 
baseline $L=1.04$ km is shown with dashed lines.}
\label{fig:f1}
\end{center}
\end{figure}

In Fig. \ref{fig:f1} we show  the effective probability $P_{eff}$,
Eq.~(\ref{eq:fluxr}), as a function of neutrino energy for different values of $\Delta 
m^2_{42}$.  
For comparison we also show  the standard probability with baseline $L$. A few comments 
are in order. 

\begin{enumerate}

\item Due to the large size of the neutrino production region, (and the effect of partial decoherence at the production)  
positions of the first and higher oscillation minima 
are shifted to lower energies in comparison with
pointlike production case with baseline $L$. 
The relative shift increases with $\Delta m^2_{42}$ and decreases with energy. 

\item The depth of oscillations is suppressed due to the decoherence effect 
in comparison to the standard oscillation case. This suppression becomes 
stronger with increase of $\Delta m^2_{42}$ and therefore $\xi$, 
and decrease of energy. 

\item For characteristic neutrino energy $E_\nu = 2$ GeV 
(responsible for the observed events with energy $\sim 1$ GeV) 
and $\Delta m^2_{42} = 1$ eV$^2$  the oscillation effect in the ND is $P_{ND}\simeq 
0.97$, as compared to the average suppression, $P_{FD}=0.95$, in the FD. Therefore 
$P_{FD}/P_{ND} = 0.98$ instead of  $0.95$ without oscillations in ND.   
So, clearly for this $\Delta m^2_{42}$ the bound on mixing 
parameter should be substantially weaker.   
For $\Delta m^2_{42} = 0.5$ eV$^2$ the oscillation effect is 0.008 
and the corresponding ratio, $P_{FD}/P_{ND} = 0.96$, is close to 
the result which MINOS uses. 
\end{enumerate}

\section{Coherence loss at the production}

In this section, we find the oscillation effect assuming that neutrinos are coherently emitted 
along the  pion trajectory. We will integrate
the amplitude of oscillation over the production region and then compute the probability of the process.

But before that, let us elaborate more on the decoherence parameter $\xi$. 
According to the exact energy-momentum conservation  in the decay process the difference 
of energies of the neutrino mass eigenstates equals in the pion rest frame 
\be
\Delta E_{42}^0 = \frac{\Delta m_{42}^2}{2 m_\pi} . 
\ee
This  difference is comparable with  the width 
\be
\Delta E_{42}^0  \sim \Gamma_0
\ee
at $\Delta m_{42}^2 \sim 7$ eV$^{2}$. 
The ratio of the two parameters,  
\be
\frac{\Delta E_{42}^0}{\Gamma_0} = 
\frac{\Delta m_{42}^2}{2 m_\pi \Gamma_0} = \alpha \xi , 
\ee
coincides with the coherence parameter up to factor $\alpha \sim 0.45$. 
Notice that the difference of momenta of the eigenstates is even bigger: 
\be
\Delta p_{42}^0 = - \frac{\Delta m_{42}^2}{2 m_\pi} \cdot 
\frac{m_\pi^2 + m_\mu^2}{m_\pi^2 - m_\mu^2}~. 
\ee
Numerically, 
\be
\xi \approx 0.33 \left(\frac{\Delta m_{42}^2}{1 {\rm eV}^2} \right). 
\ee
For $\xi \gg 1$ the energy splitting is larger than the uncertainty 
in energy, and therefore the neutrino mass eigenstates are produced 
incoherently. 

This consideration practically does not depend on the reference frame since 
violation of coherence is a physical phenomenon. 
In a  frame where the pion moves  with the Lorentz factor $\gamma_\pi$ and neutrinos  
are emitted in  the  direction of  pion we have $\Gamma_0 \rightarrow 2\gamma_\pi 
\Gamma_0$, and 
\be
\Delta E_{42}^0 \rightarrow 2\gamma_\pi \Delta 
E_{42}^0\,\frac{m_\pi^2}{m_\pi^2-m_\mu^2}, 
\ee
so that the ratio changes roughly by a factor of 2~\cite{Farzan}.

If $\xi \ll 1$ the mass eigenstates are produced in the coherent state. 
In the MINOS case we deal with the intermediate situation when results  depend 
on the shape of the wave packet of neutrino.  
Indeed, the expression  in Eq.~(\ref{eq:flux2}) depends on the exponential 
decay factor, which turns out to be 
the shape factor of the neutrino wave packet (squared) from free pion decay. 
 
In a reference frame where the pion has energy $E_\pi$ 
the decay length  equals 
$l_{dec} = E_\pi /(m_{\pi} \Gamma_0) = E_\nu / (\alpha m_{\pi} \Gamma_0)$, 
and the oscillation length of neutrinos $l_\nu = 4\pi E_\nu / \Delta m^2_{42}$.  
The ratio of the two lengths
\be 
\frac{l_{dec}}{l_\nu} = \frac{\Delta m^2}{4 \pi \alpha m_\pi \Gamma_0} 
= \frac{\xi}{2\pi} 
\ee
equals up to the factor $1/2\pi$ the decoherence parameter. Therefore 
the condition of coherence is $l_{dec} \ll l_\nu$. 

Consistent treatment of decoherence effects can be done in terms of wave 
packets. The produced $\nu_\mu$ state can be described as 
\be 
|\nu_\mu (x ,t)\rangle = \sum_{i} U_{\mu i}^* \Psi_i (x, t) |\nu_i\rangle 
= \sum_{i} U_{\mu i}^* 
g_i^S (x - v_i t) e^{i p_i x - i E_i t}|\nu_i\rangle  ,
\label{eq:prod}
\ee
and for the detected state we have 
\be
|\nu_\mu (L - x)\rangle = \sum_{k} U_{\mu k}^*
g_k^D (x - L) e^{i p_k^{\prime}(x - L)}|\nu_k\rangle. 
\label{eq:det}
\ee 
Here, $\Psi_i (x, t)$  is the wave function of the mass eigenstate and $\nu_i$,   $g_i^S$ and $g_i^D$ are the shape factors of the neutrino wave packets  which correspond to the production and detection of the $i$th mass eigenstate. Also, $p_i$ and $p^{\prime}_i$ are the mean momenta of the these wave packets,  $E_i = \sqrt{p_i^2 + m_i^2}$  and  $v_i$ is the group velocity of $i$th mass eigenstate (see \cite{akhm} for details). 

We now show that under certain assumptions  the wave packet calculation leads to the same result as  in Sec. 2. In the  case of neutrinos produced in  pion decay the neutrino wave function in the  space-time point $(x, t)$, as well as the shape factor  $g_i^S$, can be found in the  following way. The neutrino wave packet is formed as a result of an integration over  the production region of partial plane waves emitted from each point of the region: 
\be
\Psi_i(x,\,t) = \int dx_S \int dt_S 
e^{-\frac{1}{2} \Gamma t_S} e^{i p_i (x - x_S) - i E_i (t - t_S)}. 
\label{eq:psi}
\ee
Here $e^{-\frac{1}{2} \Gamma t_S}$ gives the amplitude of probability to 
find the pion at the moment of time  $t_S$.  
Due to strong interactions the pion  is produced in a small spacetime region around $x = 
0$, $t = 0$ with very short wave packets. 
The production region is thus approximately described by a delta function
\be
\delta(x_S - v_\pi t_S).  
\label{delta1}
\ee
Furthermore, because the baseline  is small for the ND, the spread of the wave packets due 
to the difference in group velocities is negligible.  Therefore, the shape 
of the neutrino wave packet (exponential in this case) is conserved. 
 The conservation of the shape implies in turn that there is a 
one-to-one correspondence between the points 
of the wave packet  in a given point $(x, t)$  and in    
point $(x_S, t_S)$ at the production.  
This correspondence is expressed as 
\be
t \simeq \frac{x_S}{v_\pi} + \frac{x - x_S}{v_\nu}.   
\label{eq:tts}
\ee
Here we take into account that before neutrino appears in the
point $x, t$  pion travels distance $x_S$ from the origin
with velocity $v_\pi$, and then neutrino travels distance $x - x_S$
with average group velocity $v_\nu$. 

From Eqs.~(\ref{eq:tts}) and (\ref{delta1})  we obtain 
\be
x_S(x, t) = \frac{v_\pi (v_\nu  t - x)}{v_\nu - v_\pi}, ~~~~ 
t_S(x, t) = \frac{x_S}{v_\pi} = \frac{(v_\nu t - x)}{v_\nu - v_\pi},  
\label{eq:conn}
\ee
and the second equality leads to another delta function for $t_S$: 
\be
\delta\left( t_S - \frac{v_\nu t-x}{v_\nu-v_\pi} \right).  
\label{delta2}
\ee
The two deltas, Eqs.~\eqref{delta1} and \eqref{delta2}, 
remove the integration over $x_S$ and $t_S$ in (\ref{eq:psi}) giving
\be
\Psi_i = 
e^{-\frac{1}{2} \Gamma t_S} e^{i p_i (x - x_S) - i E_i (t - t_S)} ,
\label{eq:prod2}
\ee
where $x_S$ and $t_S$ are now functions of $x$ and $t$ as in Eq.~(\ref{eq:conn}). 
Therefore, according to Eq.~(\ref{eq:prod}), the shape factor equals
\be
g_i^S (x - v_\nu t) = e^{-\frac{1}{2} \Gamma t_S} e^{- i p_i x_S + i E_i t_S}, ~~~
t_S = t_S(x, t) \propto (v_\nu t - x).  
\label{eq:shape}
\ee
The amplitude of probability of the $\nu_\mu$ detection is given by  
\bea
A_{\mu \mu} (t) & = &  
\int dx \langle \nu_\mu (L - x) |\nu_\mu (x ,t)\rangle \\
& = &  \sum_{i} |U_{\mu i}|^2  \int dx
g_i^S(x - v_i t) g_i^{D*}(x - L) e^{i p_i L - i E_i t}  e^{i(p_i - p_i^{\prime})(x - L)}.
\eea
Since  the size of the ND is much smaller than the decay pipe (production region) 
we can take   $g_i^{D}(x - L) \propto \delta(x - L)$ and therefore 
\be
A_{\mu \mu} (t)  =   \sum_{i} |U_{\mu i}|^2 
g_i^S(L - v_i t) e^{i p_i L - i E_i t} .
\ee
The probability equals
\be
P = \int_{-\infty}^{+\infty} dt|A_{\mu\mu}|^2 =  \int_{-\infty}^{+ \infty} dt \sum_{i,k} 
g^S_i (L - v_i t) g^{S*}_k (L - v_k t)
|U_{\mu i}|^2 |U_{\mu k}|^2 e^{i\phi_{ik}^0(L, t)}, 
\label{eq:prpr}
\ee
where $\phi_{ik}^0(L, t) \equiv (p_i - p_k)L - (E_i - E_k)t \equiv \Delta p L - \Delta E t$ .  
Using the expressions for the shape factors in Eq.~(\ref{eq:shape}) we obtain 
\be
P =  \int_{-\infty}^{+ \infty} dt e^{-\Gamma t_S (L, t)} 
\sum_{i,k} |U_{\mu i}|^2 |U_{\mu k}|^2 e^{i\phi_{ik}(L, t)},
\label{eq:prpr0}
\ee
where 
\be
\phi_{ik}(L, t) = \Delta p \big[L - v_\pi t_S(L, t)\big] - \Delta E \big[t - t_S (L, t)\big]
\label{eq:phasetot}
\ee
is the total phase difference between  the mass eigenstates arriving at the detector in the 
moment of time $t$. The time factor  in the last term of Eq.~\eqref{eq:phasetot} can be 
rewritten as  
$$
t - t_S(L, t) = \frac{L - v_\pi t_S}{v_\nu}.  
$$
Then using the relation 
\be 
\Delta E \approx v_\nu \Delta p + \frac{1}{2E} \Delta m^2
\ee 
(which follows from the dispersion relation) we obtain 
\be
\phi_{ik}(L, t) = \frac{\Delta m^2}{2E} (L -  v_\pi t_S)
\label{eq:phasetot2}
\ee
which exactly coincides with the standard oscillation phase.  Inserting this expression into 
Eq.~(\ref{eq:prpr0}) and changing the integration variable  $t$  to $x \equiv x_S = v_\pi 
t_S$, which varies in the limits $0 - l_p$, we find 
\be
P \propto  \int_0^{l_p} dx  e^{-\Gamma x} P_{\mu \mu} (L - x).
\label{eq:prpr2}
\ee
This  coincides (after normalization) with  the expression  found in  Eq.~(\ref{eq:flux2}). 
Thus, the result for completely coherent neutrino emission   along the whole neutrino 
trajectory coincides with the case of completely incoherent  emission which corresponds to 
very short wave packet of neutrino. Essentially,  here we have a situation that resembles a 
coordinate-space version
of the Kiers-Nussinov-Weiss  theorem~\cite{Kiers:1995zj} where a fully incoherent ensemble of 
neutrino states is physically  indistinguishable from neutrinos produced coherently if both 
cases lead to the same energy distribution.

If pions collide, the shape factor of the  neutrino wave packet is not determined by the decay 
function of pion \cite{prep}.  The coherent  emission will occur only between two consecutive 
collisions with substantial momentum transfer.  This usually leads to the Gaussian shape 
factor for the neutrino wave packet with the  width $\sigma_x$ determined by mean free path 
of the pion. The consideration is simple if  $\sigma_x$ is much smaller than the decay 
length: $\sigma_x \ll l_{decay}$. Suppose  the pion emits a neutrino  coherently in some 
region of size $\sigma_x$ centered at $x_S$.  Then, instead of leading to 
Eq.~(\ref{eq:pmumu}), the integration over  the emission region gives for the oscillation 
probability
\be 
P_{\mu\mu} (E_\nu, L - x) = \bar{P} +
S_{loc}(\sigma_x/l_\nu )\frac{1}{2} \sin^2 2 \theta_{24} \cos [\xi \Gamma (L - x_S)],
\label{eq:pmumu1}
\ee
where $S_{loc}(\sigma_x/l_\nu)$ is the  localization factor. With collisions, it is this 
probability that should be used in  computations of the neutrino flux in Eq.~(\ref{eq:flux2}) 
and the integration should be performed  over $x_S$.  If $\sigma_x$ does not depend on $x_S$, 
in the final expression Eq.~(\ref{eq:fluxr})  $S_{loc}(\sigma_x/l_\nu)$ appears as an 
additional factor at the oscillatory term.  For $\sigma_x \ll l_\nu$ one has $S_{loc} \approx 
1$ and Eq.~(\ref{eq:pmumu1}) reproduces the  original result. Another way to take into 
account the localization of pions is to  perfom the averaging of the oscillatory term in 
Eq.~(\ref{eq:fluxr}) over the interval $\sigma_x$.

In the case $\sigma_x  \sim l_{decay}$ 
one expects some interplay between exponential decay and decoherence 
due to finite $\sigma_x$ \cite{prep}.  In this case in the limit $\sigma_x \gg l_{decay}$ one 
should obtain again our original result.

\section{MINOS data: Bounds and hint}

The number of the NC events with a given total energy of  hadrons, $E_h$, 
equals 
\be
N(E_h) = \int dE_\nu  dx F_\nu (E_\nu) \frac{d\sigma (E_\nu, E_h)}{dx dy},
\label{eq:nev}
\ee
where $\sigma (E_\nu, E_h)$ is the cross-section  of production of hadrons with energy $E_h$ 
by neutrinos of energy $E_\nu$, $y \equiv E_h/E_\nu$  is the inelasticity and $x$ is the 
Bjorken 
variable. In what follows we will identify  $E_h$  with the reconstructed energy $E_{reco}$ 
of \cite{Adamson:2011ku}. 
We assume that $\sigma (E_\nu, E_h)$ is well approximated  by the deep inelastic scattering cross-section of 
neutrinos off nucleons. We use the MSTW 2008 set for the  parton distribution functions and the reported spectrum 
for the NuMI beam (see \cite{Kopp:2007zz} for instance).

\begin{figure}[ht]
\begin{center}
\vskip 1cm
\includegraphics[width=14cm]{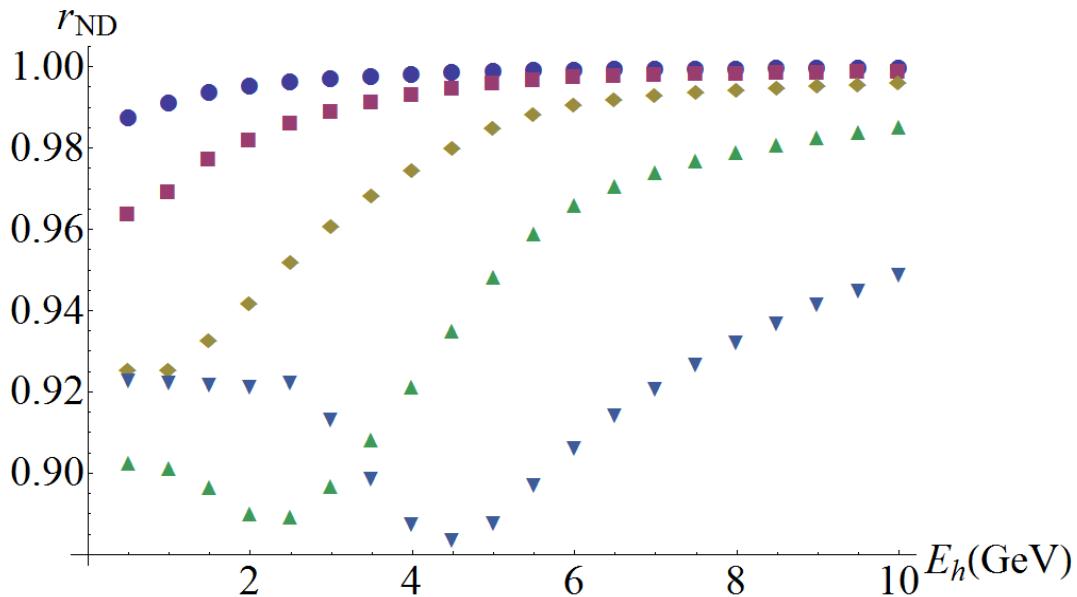} 
\end{center}
\caption{The ratio $r_{ND}$ of events at the ND with and without oscillations 
as a function of the deposited energy for $\Delta m^2_{42} = 0.5$ $\trm{eV}^2$(circles), 1 $\trm{eV}^2$(squares), 2 $\trm{eV}^2$(rhombuses), 4 $\trm{eV}^2$(upside triangles), 8 $\trm{eV}^2$(downside triangles). We take $\sin^2 2\theta_{24} = 0.1$.}
\label{fig:f2}
\end{figure}

Let $N^0(E_h)$ be the theoretical number of events  computed according to Eq.~(\ref{eq:nev}) 
for $F_\nu =  F_\nu^0$, \emph{i.e.}, as if there were  no oscillations. 
We now define the ratio of events in the ND, with and without oscillations:
\be
r_{ND} \equiv  \frac{N(E_h)}{N^0(E_h)} \,.
\ee
The ratio $r_{ND}$ as a function of the hadron (``reconstructed'') energy is shown in 
Fig.~\ref{fig:f2}.
Its behaviour with varying $\Delta m_{42}^2$  can be separated in three regions:

\begin{enumerate}

\item For $\Delta m_{42}^2 < 2\trm{ eV}^2$ the peak of the spectrum  of the incident 
neutrinos, $E_\nu \sim 3.3$ GeV, is in the region of small oscillation effect, (see Fig.~\ref{fig:f1}), 
and  $r_{ND} \sim 1$ for most of the $E_h$ range. The oscillation minimum  is below 
$E_\nu \sim  2$  GeV 
and, therefore, there is some depletion of events with energy $E_h < 2$ GeV.  Nevertheless, 
$r_{ND}$ remains bigger than $\bar{P}$ for the whole energy range.

\item For $2 \trm{ eV}^2 < \Delta m_{42}^2 < 8 \trm{ eV}^2$  the dip in the oscillation 
probability falls in the region of energies for which  the neutrino beam has maximum 
intensity. There is a considerable depletion of events  for neutrino energies below $5$ GeV 
and  $r_{ND} < \bar{P}$ in this region. We will see below that, in this case, \emph{an excess of events should be observed in the FD} compared to MINOS simulations.

\item For $\Delta m^2_{42} > 8 \trm{ eV}^2$ the first oscillation minimum  is above the peak of the neutrino spectrum  and the averaging is substantial  for low $E_h$.  Still, for $\Delta m^2_{42} < 15 \trm{ eV}^2$ there are noticeable deviation from $\bar{P}$ below $E_h = 5$ GeV as MINOS becomes sensitive to higher oscillation minima.

\end{enumerate}

\begin{figure}[t]
\begin{center}
\includegraphics[width=8cm]{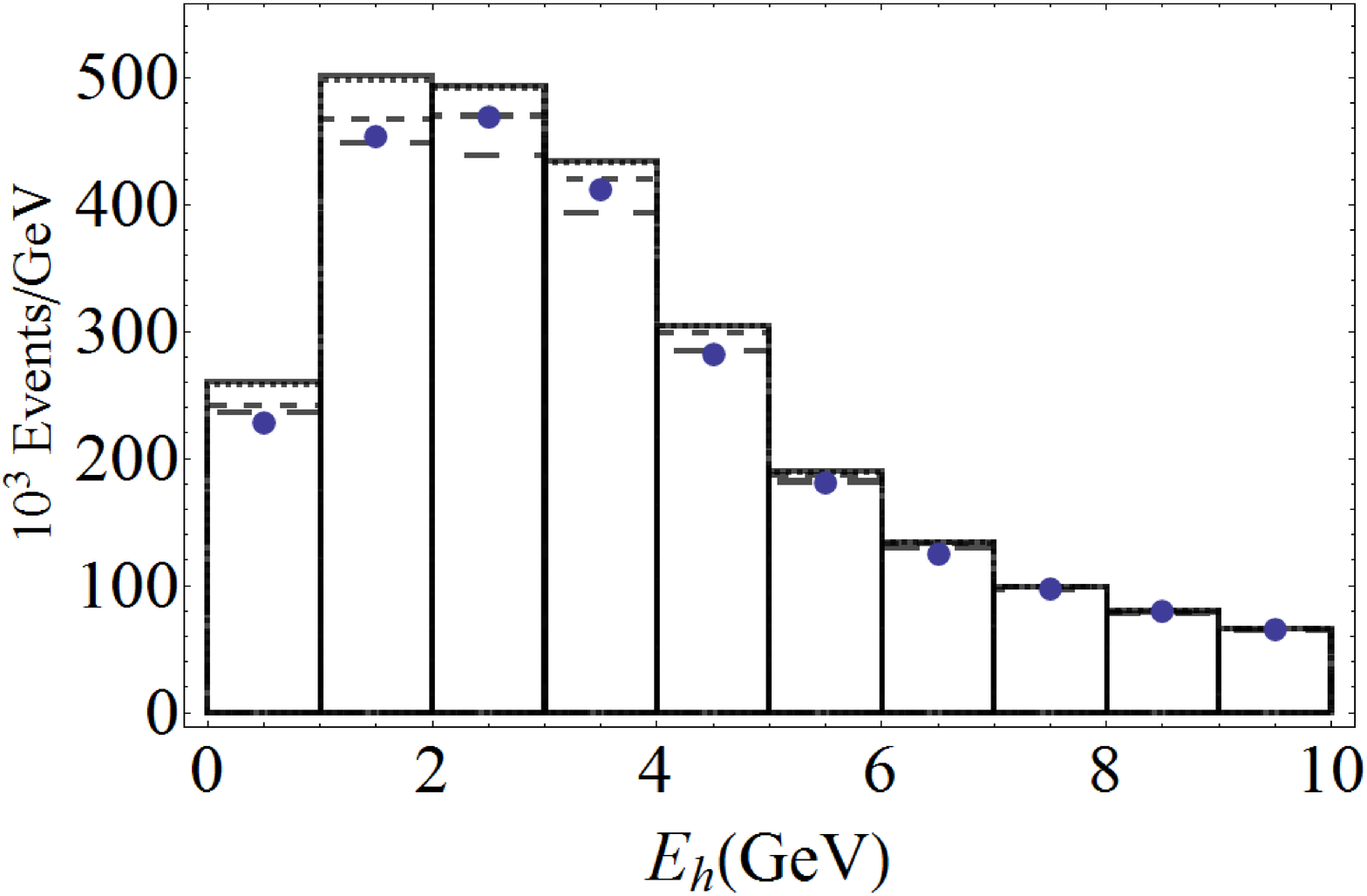} 
\includegraphics[width=8cm]{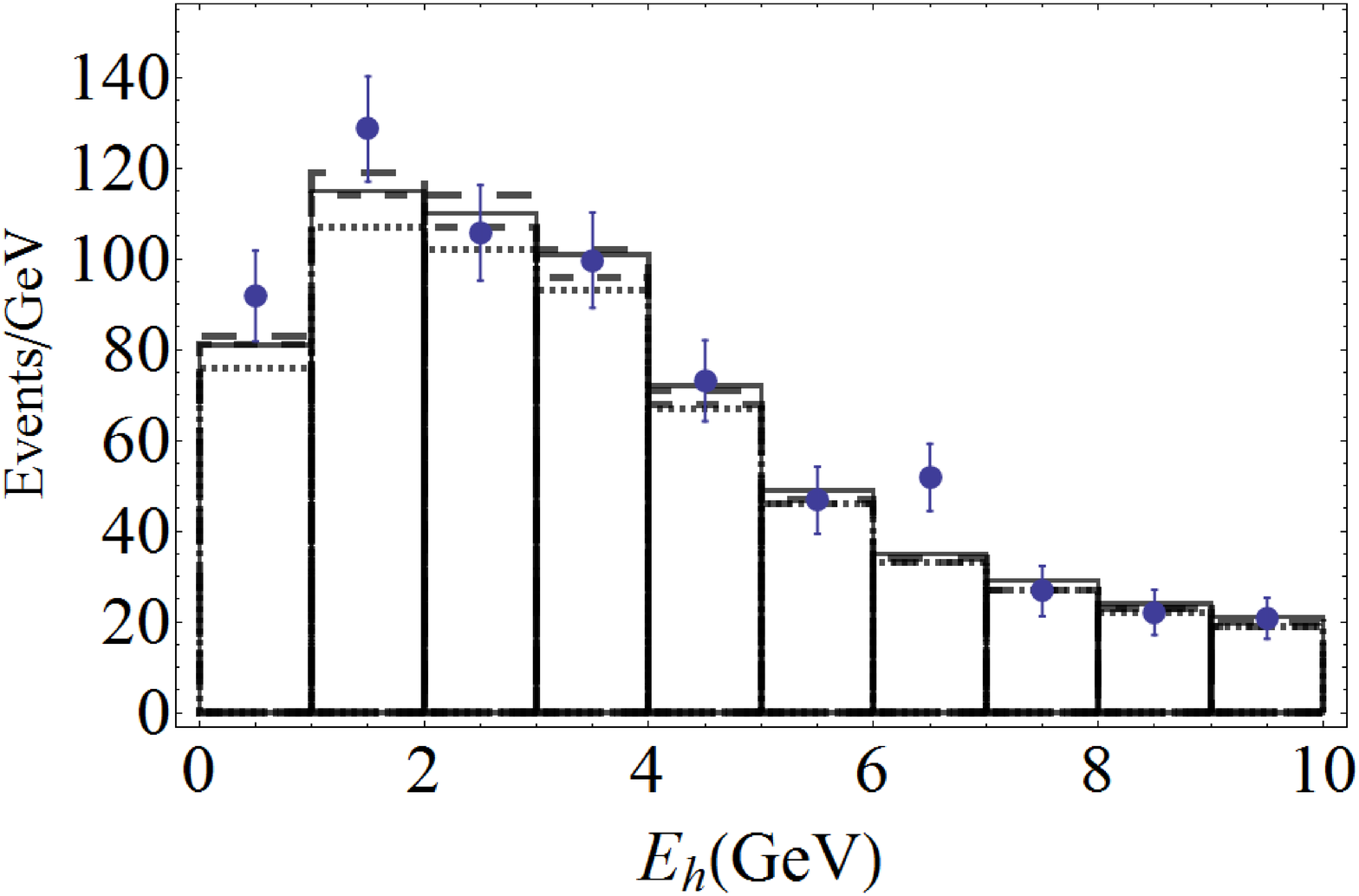}
\caption{Predictions for the number of events at the ND (left) and FD (right) with and without $\nu_\mu - \nu_s$ oscillations. Solid histograms show predictions without $\nu_\mu - \nu_s$ oscillations, $N_{ND}^0$ and $N_{FD}^0$. The other histograms correspond to the case of $\nu_\mu - \nu_s$ oscillations with  $\sin^2 2\theta_{24} = 0.15$ and different values of $\Delta m^2_{42}$: 0.5  eV$^2$ (dotted), 2  eV$^2$ (short dashed), 4 eV$^2$ (long dashed), see Eqs.~\eqref{nd0-nd} and \eqref{NFD}. The dots with the error bars represent the MINOS data points.}
\label{fig:f3}
\end{center}
\end{figure}

The Monte Carlo prediction $N_{ND}^0$ for the number of events 
in the ND without oscillations,  (see Fig. 1 in \cite{Adamson:2011ku}), needs to be corrected 
when the oscillations into sterile neutrino  are present as
\be  
N_{ND} = r_{ND}~ N_{ND}^0 \,.
\label{nd0-nd}
\ee
At 
\be
\Delta m_{42}^2 = (1 - 3)~ {\rm eV}^2, ~~~    
\sin^2 2\theta_{24} =   0.15 - 0.20 \,,
\label{mix}
\ee
the oscillations can explain the deficit of events in the ND in the low energy bins (see Fig.~\ref{fig:f3}).   
This deficit is not yet statistically significant and for   
ND  it is  below the systematic errors.  However the oscillations at ND 
must be taken into account when oscillation effect ``propagates'' from ND to FD.  

In Fig. \ref{fig:f3} we show distributions of events  over the reconstructed hadron energy $E_h$  in the ND for different values of $\Delta m^2_{42}$ (Eq. (53)).
Also shown is the distribution without oscillations $N_{ND}^0$ from \cite{Adamson:2011ku} and experimental points.

The MINOS collaboration predicts the number of events in the FD, $N_{FD}^0$,  extrapolating the experimentally  
measured spectrum at the ND. In \cite{Adamson:2011ku} it is  assumed that oscillation effects are absent in the 
ND  but the effect of the usual $3\nu$ oscillations at the FD is included  
(see Fig.~2 in~\cite{Adamson:2011ku}). The averaged $\nu_\mu - \nu_s$ 
oscillations in the FD modify this prediction: $N_{FD}^0 \rightarrow \bar{P}N^0_{FD}$.
However, when oscillations at the ND are taken  into account,  the latter must be 
corrected consistently with Eq.~(\ref{nd0-nd}):  
\be
N_{FD} = \frac{\bar{P}}{r_{ND}} ~N_{FD}^0 \,. \label{NFD}
\ee
Thus, the factor
\be
R(E_h) = \frac{\bar{P}}{r_{ND}} .
\ee
is the conversion factor for the flux predicted at the FD 
with and without of  $\nu_\mu-\nu_s$ oscillations.

The behavior of $R(E_h)$ can be straightforwardly inferred from Fig.~\ref{fig:f2}. 
Again, there are three different  cases:

\begin{enumerate}

\item Averaged effect in ND: $R(E_h) \approx 1$. This happens  at low energies for large $\Delta m_{42}^2$ when one has the averaged oscillation effect in the ND. The MINOS prediction for the number of events at the FD is unchanged by $\nu_\mu - \nu_s$ oscillations. 

\item Strong suppression in ND: $R(E_h) > 1 $.   
The neutrino energies in  the peak of the neutrino flux are  around 
the first oscillation minimum for  $2 \trm{ eV}^2 < \Delta m_{42}^2 < 8 \trm{ eV}^2$. 
In this case one predicts  some excess of events in the FD in comparison to 
the MINOS extrapolation which can partially explain the observed excess 
(see Fig.~\ref{fig:f3}) 

\item Weak effect in the ND: $R(E_h)< 1$. This regime is realized for neutrino energies above the 
first oscillation minimum for all mass squared differences.   It corresponds to small or no 
oscillation effect at the ND. 

\end{enumerate}

In Fig~\ref{fig:f3} we show the distribution  of events
in the reconstructed hadron energy,
$E_h$  in the FD for different values of $\Delta m^2_{42}$ (Eq. (55))
Also shown is the distribution without oscillations $N_{ND}^0$ from \cite{Adamson:2011ku}
and experimental points.


We consider next how the bounds on  the $\nu_\mu - \nu_s$ mixing are modified for different  $\Delta 
m_{42}^2$.
For $\Delta m_{42}^2 < 0.5$ eV$^2$  the decoherence  is negligible, the oscillation 
effects in ND are significant only  at low energies and therefore the MINOS limit in 
Eq.~(\ref{limit}) is approximately valid. 
For  $\Delta m_{42}^2 > 0.5$ eV$^2$ the limit should be modified. 
The strongest modification is for $\Delta m_{42}^2 =  (2 - 8)$ eV$^2$, 
when the first oscillation 
minimum is at neutrino energies $E_\nu$ that correspond to  
the peak in the spectrum of events. 
Notice that Eq.~(\ref{mix}) corresponds  to $\theta_{24} \sim (11 - 13)^{\circ}$ which is 
substantially larger than the limit in Eq.~(\ref{limit}).

We stress that the calculations performed in this section 
are meant for illustration purposes. For precise quantitative 
results one should perform a complete MC simulation 
of events without the simplifications we made.

\section{Conclusions}

 

In this work we considered the oscillation effects in the ND of the  MINOS 
experiment. The MINOS setup  realizes an  interesting situation of partial 
decoherence of the neutrino state at the production, when the energy splitting  
of two mass eigenstates  is comparable with the  energy uncertainty 
of the initial state (the width of pion). Decoherence leads to suppression of the oscillation effect at high energies (above the oscillation minimum), 
to the shift of oscillation minimum (dip)  to low energies 
and to suppression of the depth of oscillations. The 
suppression becomes stronger at low energies.  In general, the effect of decoherence should be taken into 
account for all experiments that perform searches for sterile neutrinos with 1 eV 
mass using neutrino beams from pion  decays. 

The MINOS bounds, Eq.~(\ref{limit}) remain valid for 
$\Delta m^2_{42} <0.5$ eV$^2$.   For $\Delta m^2_{42} > 0.5$ eV$^2$ 
the oscillation effect in the ND should be taken into 
account. It should noticeably weaken the bounds for  
$\Delta m^2_{42}  > 2$ eV$^2$. An estimation of this effect has been 
performed in this paper; a full quantitative analysis  
should be done by the MINOS collaboration. For $\Delta m^2_{42} > 15$ eV$^2$ the 
bound disappears and MINOS is insensitive to the $\nu_\mu - \nu_s$ oscillations.

The MINOS data might actually provide  a hint of oscillations 
with $\Delta m^2_{42} \sim (1 - 3)$ eV$^2$. The 
oscillations can explain some deficit of signal 
in  the low energy bins of the  ND as compared to the Monte Carlo 
prediction, and the excess of events in the FD. 


\section*{Acknowledgment} 
The authors are grateful to E. Kh. Akhmedov for very useful discussions. 

\section*{Appendix. Relation between the neutrino and pion energies}

Using the Lorentz transformation to the laboratory frame 
we have for neutrino momenta:  
\be
p_y = p_0 \sin \theta_0, ~~~p_x \approx p_0 \gamma_\pi  (\cos \theta_0 + \beta_\pi), 
\label{eq:momenta}
\ee
where $\theta_0$ is the neutrino angle of emission in the pion rest frame; 
$\gamma_\pi \equiv E_\pi / m_\pi$ is the Lorentz factor of the pion  and 
$\beta_\pi$ is the pion velocity. The neutrino energy in the laboratory frame,  
$E_\nu = \sqrt{p_x^2 + p_y^2}$, in the aproximation  of $\gamma_\pi \gg 1$ can then be 
written as 
\be 
E_\nu \approx  E_0 \gamma_\pi  (\cos \theta_0 + \beta_\pi) 
\approx E_0 \gamma_\pi (\cos \theta_0 + 1).  
\label{eq:enu}
\ee
Therefore,
\be
\alpha \approx  \frac{E_0}{m_\pi} (\langle \cos \theta_0 \rangle + 1) . 
\ee
To evaluate the effective value of the angle $\langle \cos \theta_0 \rangle $
let us consider the real experimental setup.  
According to Eq.~(\ref{eq:momenta}), the angle between the neutrino and pion in the 
laboratory frame  equals
\be
\tan \theta_\nu = \frac{\sin \theta_0}{\gamma_\pi (\cos \theta_0  + 1)}. 
\ee
Consequently
\be
\cos \theta_0 = \frac{1 - b}{1 + b}, ~~~b \equiv \gamma_\pi^2 \tan^2 \theta_\nu. 
\ee
The maximal value of $\tan \theta_\nu$ is given by 
\be
\tan \theta^{max} = \frac{r}{L - x}
\ee
with $r \approx 1$ m  being the radius of the fiducial zone 
of the ND
and $L$ is the distance from the target to the ND. 
It equals $2.5 \times 10^{-3}$ for $x = l_p$ and 
and $10^{-3}$ for $x = L$. So, for the typical pion energy $E_\pi = 10$ GeV we find 
the  effective distance $\bar{x} = 300$ m  and  $b = 0.01$, and therefore $\cos \theta_0 
\sim  0.98$. We have checked that results  do not change substantially for different 
values of $\alpha$.


\end{document}